\documentclass{PoS}
\usepackage{amsmath,amssymb}

\title{Construction of the $\eta\to3\pi$ (and $K\to3\pi$) amplitudes using dispersive approach}
\ShortTitle{Construction of the $\eta\to3\pi$ (and $K\to3\pi$) amplitudes using dispersive approach}

\author{\speaker{Martin Zdrahal}\,\thanks{
This work was supported in part by the Center for Particle Physics
         (project no.\ LC 527), GACR (grant no.\ 202/07/P249) and by the EU Contract
         No.\ MRTN-CT-2006-035482, \lq\lq{\sc Flavia\it net}''.
         We are grateful to J.~Bijnens for providing us with the Fortran code that generated $\chi$PT $\eta\to3\pi$ results used in this proceedings.}
         $^{\,ab}$, Karol Kampf$^{\,*}$\thanks{Till 30 September 2009 the affiliation was Laboratory of Particle Physics, PSI, Switzerland.}$\ {}^{\,bc}$, Marc Knecht$^{\,d}$ and Ji\v r\'i Novotn\'y$^{\,b}$\\
$^a$ Faculty of Physics, University of Vienna, Boltzmanngasse 5, A-1090 Vienna, Austria\\
$^b$ Charles University, Faculty of Mathematics and Physics, IPNP, V Hole\v{s}ovi\v{c}k\'ach 2, CZ-180 00 Prague, Czech Republic\\
$^c$ Department of Theoretical Physics, Lund University, S\"olvegatan 14A, SE 223-62 Lund, Sweden\\
$^d$ Centre de Physique Th\'eorique\thanks{Unit\'e Mixte de Recherche  (UMR 6207) du CNRS  et des Universit\'es Aix--Marseille 1,
Aix--Marseille 2 et du Sud Toulon--Var, Laboratoire affili\'e \`a la FRUMAM (FR 2291)},\, CNRS-Luminy, Case 907, F-13288 Marseille Cedex 9, France\\
E-mail: \email{zdrahal@ipnp.troja.mff.cuni.cz} and\/ \email{karol.kampf@thep.lu.se}}
\abstract{The dispersive approach based only on very general principles, unitarity, analyticity and crossing symmetry, combined with chiral counting
enables to construct fully relativistic model-independent representations of the $\eta$ and $K$ three-pion decays valid up to and including two-loop corrections.
Here we demonstrate the procedure on the $\eta\to3\pi^0$ decay. We perform possible matchings of our dispersive approach with the two-loop $\chi$PT result
and briefly discuss corresponding predictions of the Dalitz plot slope parameter $\alpha$.}

\FullConference{6th International Workshop on Chiral Dynamics, CD09\\
        July 6-10, 2009\\
        Bern, Switzerland}

\renewcommand{\logo}
{\raisebox{-5mm}
{\parbox{\textwidth}{\begin{flushright}
CPT-P114-2009\\
UWThPh-2009-12
\end{flushright}
\includegraphics{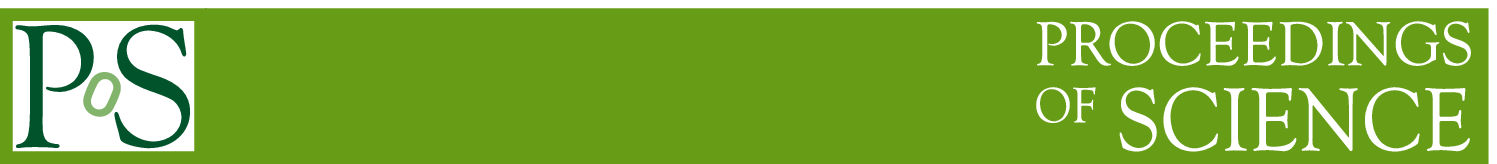}}
}}

\begin{document}

\section{Introduction}
The $K\rightarrow3\pi$ and the $\eta\rightarrow3\pi$ decays are important sources of information for the studies of
isospin breaking effects. From the branching ratio of the $\eta\rightarrow3\pi$ decay, which is forbidden in the
isospin limit, we can obtain values of the isospin breaking parameters like $R=\frac{m_s - {\widehat m}}{m_d-m_u}$.
The final state interactions of two neutral pions coming from the $K\rightarrow3\pi$ or the $\eta\rightarrow3\pi$ decays
resulting in the so-called cusp effect offer quite a simple experimental determination of the $S$-wave $\pi\pi$
scattering lengths. This has been already successfully done for the $K^+\rightarrow\pi^+\pi^0\pi^0$ decay and nowadays
there appear a large effort in a careful measurement of the cusp effect also in the case of $K_L\to3\pi^0$ and $\eta\to3\pi^0$
decays.

These are just few of the reasons why these decays attract so much attention and are still under intensive studies, both from the
experimental and from the theoretical points of view. For the sake of limited space here, we refer the reader for a more comprehensive list of these works to our
previous proceedings \cite{nase proceedings}. Nevertheless, for the reader's convenience we list here the experimental collaborations
undertaking these measurements \cite{kolaborace} and theoretical studies dealing with $\eta\to3\pi$ decay \cite{BG}--\cite{eta theory}
as we concentrate here mainly on results of this decay.

The processes in question are (generically denoted as $P\to\pi\pi\pi$ in the following section)
\begin{equation}\label{procesy}
K^+\rightarrow\pi^+\pi^0\pi^0,\ \pi^+\pi^+\pi^-;\ \ K_L\rightarrow\pi^0\pi^0\pi^0,\ \pi^0\pi^+\pi^-;\ \
\eta\rightarrow\pi^0\pi^0\pi^0,\ \pi^0\pi^+\pi^-; \ \ K_S \rightarrow\pi^0\pi^+\pi^- .
\end{equation}

The standardly used model-independent parametrization of the energetic dependence of the Dalitz plot for these decays
(Dalitz parametrization of PDG) does not account for the analytic structure arising from final state interactions
in the presence of isospin breaking.

\section{Parametrization and reconstruction procedure}
We propose a unified model-independent parametrization of the corresponding amplitudes valid at two loops relying
on very general principles, unitarity, analyticity and crossing symmetry, combined with chiral counting.
Our treatment is fully relativistic and the $\pi\pi$ scattering parameters appearing there can be chosen so
that instead of their chiral expansion, we use their physical interpretation (values) up to (and including) two-loop order.

The parametrization takes the form
\begin{equation}\label{tvar amplitudy}
\mathcal{A}(s,t,u)=\mathcal{N}_F\left[\mathcal{P}(s,t,u)+\mathcal{U}(s,t,u)\right]+O(p^8).
\end{equation}
$\mathcal{N}_F$ is an overall normalization, the polynomial part $\mathcal{P}(s,t,u)$ contains free parameters describing the energy
dependence of the processes in analogy to the Dalitz parameters and all the non-analytic part of the amplitude connected with the final
state interactions is contained in $\mathcal{U}(s,t,u)$. It depends on the parameters from the polynomial part and on parameters describing
the interaction amplitudes for the intermediate states. In the decay region and up to (and including) two loops, the only such interactions one has to take into account are the
$\pi\pi$ rescatterings -- the intermediate states other than the ones with two pseudo-Goldstone bosons are suppressed to the $O(p^8)$ order
up to a polynomial contribution included already in $\mathcal{P}(s,t,u)$. Since we are interested in the amplitude in the decay region of the Dalitz
plot (i.e.~far from the $P\pi$ threshold), the contributions of other two meson states, like $P\pi$, can be reasonably expanded up to third order
in the Mandelstam variables and included into $\mathcal{P}(s,t,u)$.

Our reconstruction procedure uses the methods of reconstruction theorem from \cite{theorem} and will be discussed into detail in \cite{budouci}.
A brief description of it can be found also in our proceedings \cite{nase proceedings}. From the theorem we obtain that the polynomial
$\mathcal{P}(s,t,u)$ is a third order polynomial in the Mandelstam variables having the same $s,t,u$ symmetries as the amplitude $\mathcal{A}(s,t,u)$,
while the unitarity part $\mathcal{U}(s,t,u)$ is determined from the single-variable dispersive integrals over the imaginary parts of S and P partial
waves of all the crossed amplitudes (obtained from the unitarity relation).

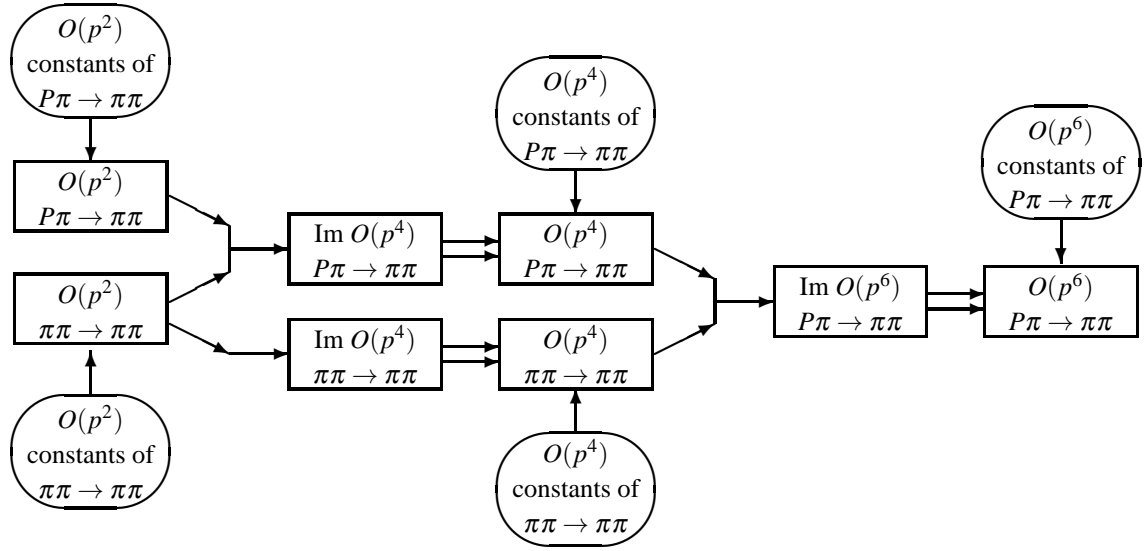
\begin{figure}
\begin{center}
\setlength{\unitlength}{1mm}%
\begin{picture}(149,69)(0,-54)
\thicklines\large
\put(0,2){\makebox(20,15){\parbox{20mm}{\small\centering $O(p^2)$ \\constants of\\ $P\pi\to\pi\pi$}}}
\put(10,9.5){\oval(21,15)}
\put(10,2){\vector( 0,-1){6}}
\put(0,-13){\framebox(20,9){\parbox{20mm}{\small\centering $O(p^2)$ \\$P\pi\to\pi\pi$}}}
\put(20.5,-8.5){\vector( 2,-1){8}}
\put(0,-50){\makebox(20,15){\parbox{20mm}{\small\centering $O(p^2)$\\ constants of\\ $\pi\pi\to\pi\pi$}}}
\put(10,-42.5){\oval(21,15)}
\put(10,-35){\vector( 0,1){6}}
\put(0,-28){\framebox(20,9){\parbox{20mm}{\small\centering $O(p^2)$ \\$\pi\pi\to\pi\pi$}}}
\put(20.5,-22.5){\vector( 2,1){8}}
\put(20.5,-25.5){\vector( 2,-1){8}}
\put(28.5,-12.5){\line(0,-1){6}}
\put(28.5,-15.5){\vector(1,0){8}}
\put(36.5,-20){\framebox(20,9){\parbox{20mm}{\small\centering Im $O(p^4)$\\ $P\pi\to\pi\pi$}}}
\put(56.5,-14.5){\vector(1,0){8}}
\put(56.5,-16.5){\vector(1,0){8}}
\put(64.5,-20){\framebox(20,9){\parbox{20mm}{\small\centering $O(p^4)$\\ $P\pi\to\pi\pi$}}}
\put(64.5,-5){\makebox(20,15){\parbox{20mm}{\small\centering $O(p^4)$\\ constants of\\ $P\pi\to\pi\pi$}}}
\put(74.5,2.5){\oval(21,15)}
\put(74.5,-5){\vector( 0,-1){6}}
\put(85,-15.5){\vector( 2,-1){8}}
\put(28.5,-29.45){\vector(1,0){8}}
\put(36.5,-34){\framebox(20,9){\parbox{20mm}{\small\centering Im $O(p^4)$\\ $\pi\pi\to\pi\pi$}}}
\put(56.5,-28.5){\vector(1,0){8}}
\put(56.5,-30.5){\vector(1,0){8}}
\put(64.5,-34){\framebox(20,9){\parbox{20mm}{\small\centering $O(p^4)$\\ $\pi\pi\to\pi\pi$}}}
\put(64.5,-51){\makebox(20,8){\parbox{20mm}{\small\centering $O(p^4)$\\ constants of\\ $\pi\pi\to\pi\pi$}}}
\put(74.5,-47.5){\oval(21,15)}
\put(74.5,-40){\vector( 0,1){6}}
\put(85,-29.5){\vector( 2,1){8}}
\put(93,-19.5){\line(0,-1){6}}
\put(93,-22.5){\vector(1,0){8}}
\put(101,-27){\framebox(20,9){\parbox{20mm}{\small\centering Im $O(p^6)$ $P\pi\to\pi\pi$}}}
\put(121,-21.5){\vector(1,0){8}}
\put(121,-23.5){\vector(1,0){8}}
\put(129,-27){\framebox(20,9){\parbox{20mm}{\small\centering $O(p^6)$ $P\pi\to\pi\pi$}}}
\put(129,-11.5){\makebox(20,15){\parbox{24mm}{\small\centering $O(p^6)$\\ constants of\\ $P\pi\to\pi\pi$}}}
\put(139,-4){\oval(21,15)}
\put(139,-11.5){\vector( 0,-1){6}}
\end{picture}
\end{center}
  \caption{Scheme of the reconstruction procedure.}
\end{figure}

A simple diagram of this procedure is given in Fig.~1. At every order a given parameter within the unitarity part (computed from the imaginary part)
corresponds exactly to the polynomial parameter of the lower order whereas the corresponding parameter within the polynomial part contains also
contributions (corrections) of the higher order. We can deal with the appearance of the parameters of these two orders within the results in two ways:

\begin{itemize}
\item fit order-by-order -- we can exactly respect the chiral orders and distinguish contributions of every order. We can further simplify the fit
by starting with fitting the lower order formula and afterwards fit only contributions of the higher orders to the polynomial part.
\item resummed fit -- we can include also part of the neglected orders by replacing all the parameters by their highest order values. This possibility
is natural in the cases where we want to give or keep the physical meaning of the parameters (this will be of use e.g.~when fitting scattering lengths
from the cusp data).
\end{itemize}

The amplitudes of all the processes (\ref{procesy}) depend on the same $\pi\pi$ scattering parameters. We can choose them to be the scattering lengths,
effective range parameters etc.~and write the $\pi\pi$ scattering amplitude in such parametrization where they keep their physical interpretation up
to two loops.

\section{Results}
At present we have a closed analytic form of the parametrization for the $K_L\to3\pi$ and $\eta\to3\pi$ decays valid up to two loops (including the pion mass differences). Nevertheless, we present here the fit of our results for $\eta\to3\pi$ decay, calculated in the first order in isospin breaking, where the interesting
physics appears already there.

We have put a part of the parametrization of this process already in our previous proceedings \cite{nase proceedings} and its explicit form will be given
in \cite{budouci}. Let us remind here that at this isospin-breaking order, the charged $\eta\to\pi^0\pi^+\pi^-$ and the neutral $\eta\to3\pi^0$ decays
are related and up to two loops we have 6 real parameters describing the energy dependence of this decay together with 4 parameters describing $\pi\pi$
scattering. We have taken the $\pi\pi$ parameters from Stern et al.~\cite{theorem} and fitted the results of Bijnens and Ghorbani \cite{BG} as follows.

\subsection{Order-by-order fit}
Our parametrization with parameters obtained by order-by-order fit should reproduce the chiral perturbation theory results \cite{BG} on the physical
decay region. This was indeed verified. On Fig.~2 and Fig.~3 we plot real and imaginary parts of the $O(p^4)$ and the $O(p^6)$ amplitudes of $\eta\to3\pi^0$
on one particular cut in kinematical variables ($t=u$). The pictures are similar also for other cuts. At every order we have taken the constants within
the unitary part from the lower order and then fitted just the polynomial part. It means that the imaginary part at each order is not fitted to the $\chi$PT
points. The $O(p^4)$ imaginary part under the $\pi\eta$ (and $KK$) threshold is reproduced exactly in the case when our amplitude is defined consistently to the chiral one (e.g.~GMO relations). The influence of the $\pi\eta$
and the $KK$ intermediate states is visible already at $O(p^4)$ above the physical region.

\begin{figure}
\begin{centering}
  \mbox{}\hfill\includegraphics[width=0.47\linewidth]{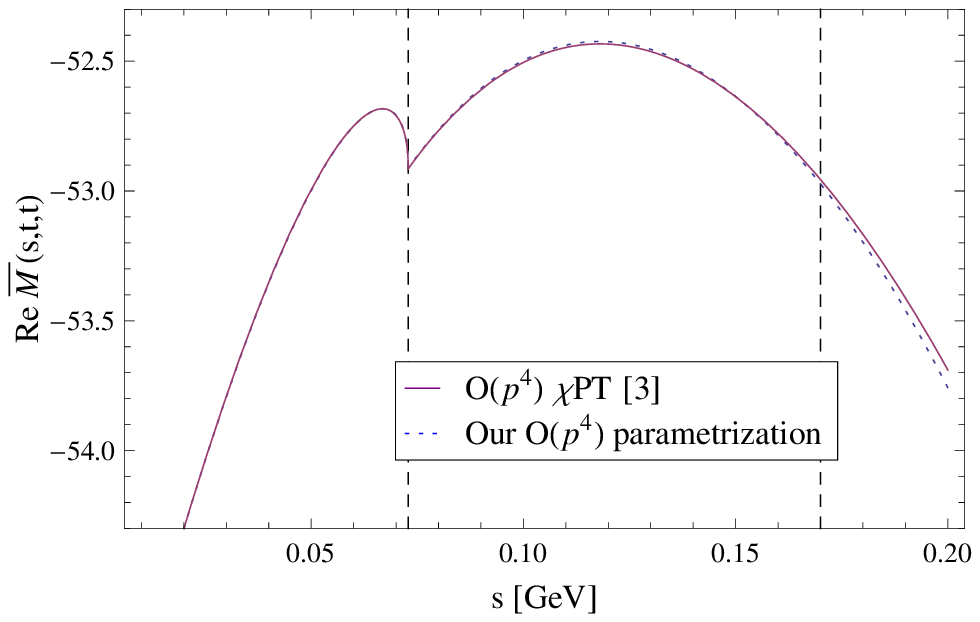}\hfill\includegraphics[width=0.47\linewidth]{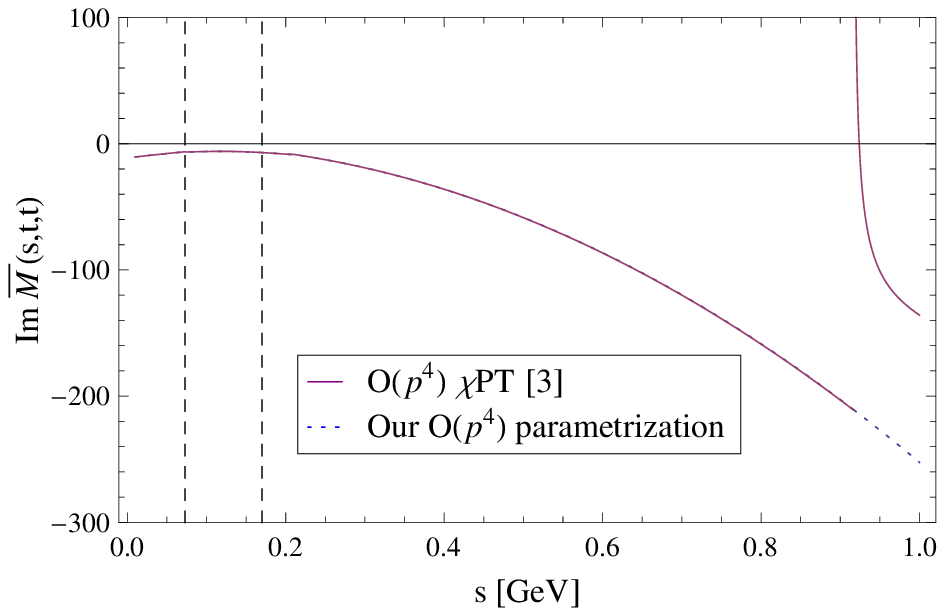}\hfill\mbox{}
  \caption{A comparison between the one-loop chiral perturbation theory result \cite{BG} and our one-loop parametrization of the amplitude
  $\eta\to3\pi^0$ on the cut $t=u$. The real part of our parametrization is fitted on the chiral results, whereas the imaginary part of it
  comes from the LO values of our constants. With the dashed lines the physical decay region is indicated.}
\end{centering}
\end{figure}
\begin{figure}
\begin{centering}
  \mbox{}\hfill\includegraphics[width=0.46\linewidth]{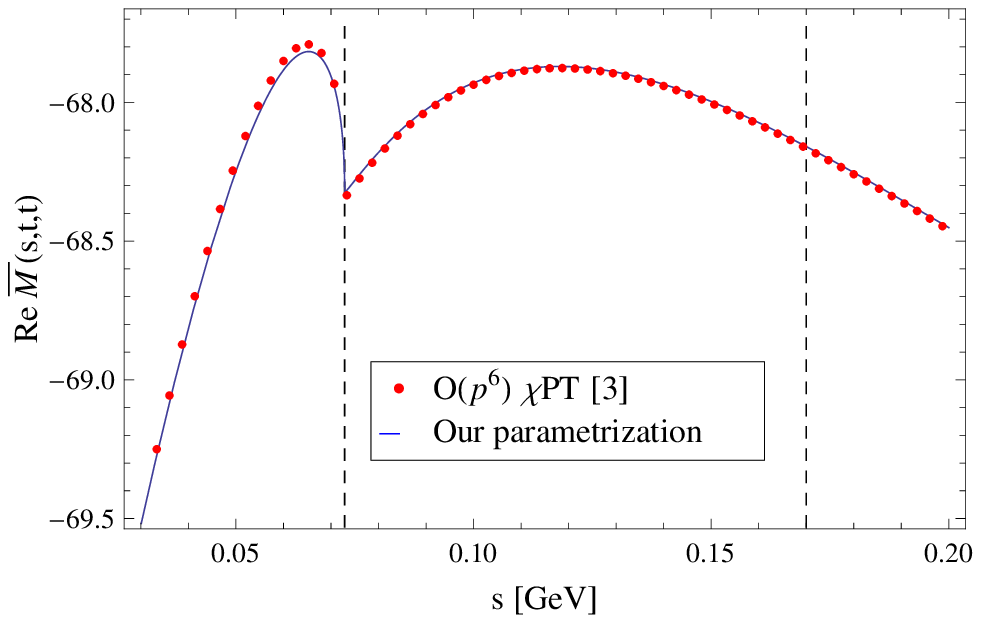}\hfill\includegraphics[width=0.46\linewidth]{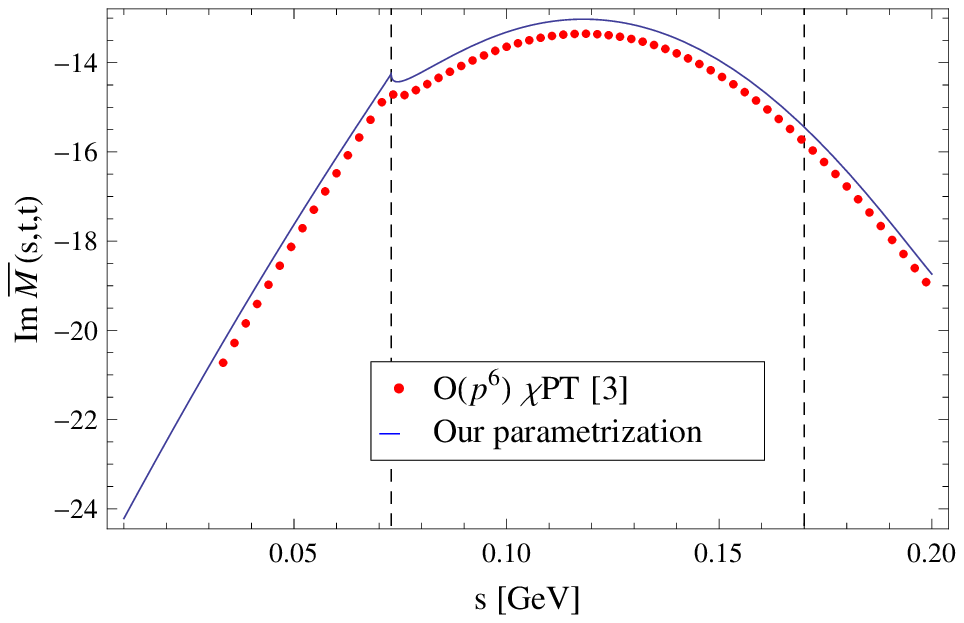}\hfill\mbox{}
  \caption{A comparison between the two-loop $\chi$PT result \cite{BG} and our parametrization of the amplitude
  $\eta\to3\pi^0$ on the cut $t=u$. The real part of our parametrization is fitted on the chiral results, whereas the imaginary part of it
  comes from the values of our constants determined from the $O(p^4)$ fit from Fig.~2.}
\end{centering}
\end{figure}
\begin{figure}
\begin{centering}
  \mbox{}\hfill  \includegraphics[width=0.46\linewidth]{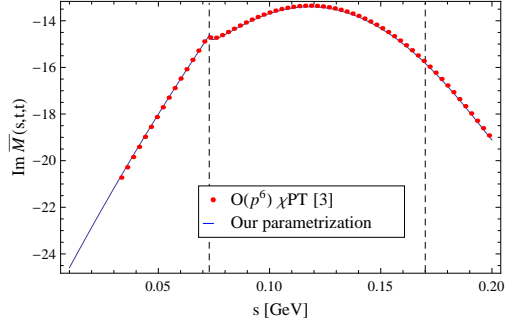}\hfill\mbox{}
  \caption{The same comparison of imaginary parts as in Fig.~3 with adding also a small imaginary part to the constant parameter in our parametrization
  that is fitted from the chiral result.}
\end{centering}
\end{figure}

The small shift in the imaginary part of $O(p^6)$ is caused by absorptive contributions from sunset diagrams. Their effect
can be included by adding a small imaginary part to the $O(p^6)$ parameters. However, as demonstrated on Fig.~4, to have a good correspondence, we need in fact
just a small constant shift which is suppressed by two orders of magnitude with respect to the real part of the $O(p^6)$ contribution
to the constant term.

\subsection{Order-by-order fit vs.~resummed fit}
The further illustration can shed light on the sign discrepancy between $\chi$PT \cite{BG} and the dispersive determination \cite{disperze} of the Dalitz parameter
$\alpha$ of the neutral $\eta\to3\pi^0$ decay. The Dalitz parametrization of this process is defined according to
\begin{equation}
\lvert\overline{M}(s,t,u)\rvert^2=\overline{A}_0^2\left(1+2\alpha z+\dots\right), \qquad z=\frac{2}{3}\sum_{i=1}^3\left(\frac{3E_i-M_\eta}{M_\eta-3M_\pi}\right)^2,
\end{equation}
where $E_i$ is the energy of the $i-$th pion in the final state. The $\chi$PT analysis \cite{BG} gives a positive value for $\alpha$
whereas the dispersive approaches from \cite{eta theory} (and also experiments) predict negative values.

\begin{figure}
\begin{centering}
  \mbox{}\hfill\resizebox{!}{0.29\linewidth}{\setlength{\unitlength}{0.0500bp}%
  \begin{picture}(6100.00,3900.00)(500,200)%
      \put(3600,3957){\makebox(0,0){\strut{}\large $\lvert \overline M(s,t,u)\rvert^2$}}%
      \put(1846,914){\makebox(0,0)[r]{\strut{} 0.05}}%
      \put(2698,791){\makebox(0,0)[r]{\strut{} 0.1}}%
      \put(3551,668){\makebox(0,0)[r]{\strut{} 0.15}}%
      \put(4402,545){\makebox(0,0)[r]{\strut{} 0.2}}%
      \put(2998,441){\makebox(0,0){\strut{}\large s [GeV]}}%
      \put(4892,783){\makebox(0,0){\strut{} 0.05}}%
      \put(5384,996){\makebox(0,0){\strut{} 0.1}}%
      \put(5876,1210){\makebox(0,0){\strut{} 0.15}}%
      \put(6368,1423){\makebox(0,0){\strut{} 0.2}}%
      \put(6029,828){\makebox(0,0){\strut{}\large t [GeV]}}%
      \put(920,1614){\makebox(0,0)[r]{\strut{} 5000}}%
      \put(920,2120){\makebox(0,0)[r]{\strut{} 5500}}%
      \put(920,2626){\makebox(0,0)[r]{\strut{} 6000}}%
    \put(0,0){\includegraphics{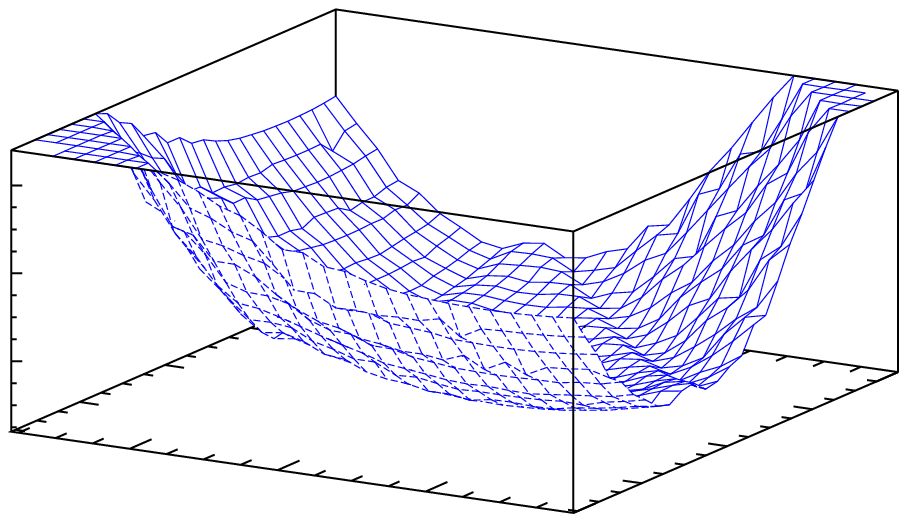}}%
  \end{picture}}\hfill
  \resizebox{!}{0.29\linewidth}{\setlength{\unitlength}{0.0500bp}%
  \begin{picture}(6100.00,3900.00)(500,200)%
      \put(3600,3957){\makebox(0,0){\strut{}\large $\lvert \overline M(s,t,u)\rvert^2$}}
      \put(1846,921){\makebox(0,0)[r]{\strut{} 0.05}}%
      \put(2698,798){\makebox(0,0)[r]{\strut{} 0.1}}%
      \put(3551,675){\makebox(0,0)[r]{\strut{} 0.15}}%
      \put(4402,552){\makebox(0,0)[r]{\strut{} 0.2}}%
      \put(2998,441){\makebox(0,0){\strut{}\large s [GeV]}}%
      \put(4892,800){\makebox(0,0){\strut{} 0.05}}%
      \put(5384,1013){\makebox(0,0){\strut{} 0.1}}%
      \put(5876,1227){\makebox(0,0){\strut{} 0.15}}%
      \put(6368,1440){\makebox(0,0){\strut{} 0.2}}%
      \put(6029,878){\makebox(0,0){\strut{}\large t [GeV]}}%
      \put(920,1208){\makebox(0,0)[r]{\strut{} 500}}%
      \put(920,1749){\makebox(0,0)[r]{\strut{} 1500}}%
      \put(920,2288){\makebox(0,0)[r]{\strut{} 2500}}%
      \put(920,2829){\makebox(0,0)[r]{\strut{} 3500}}%
    \put(0,0){\includegraphics{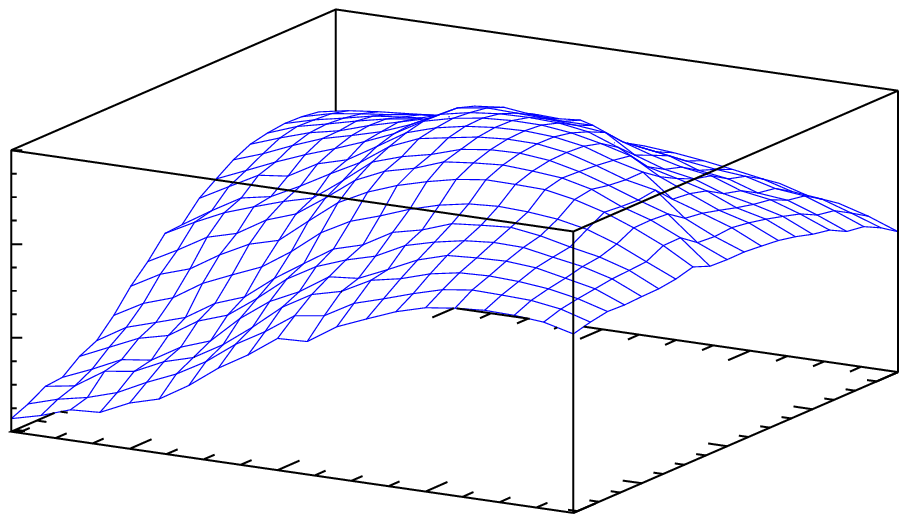}}%
  \end{picture}}%
  \hfill\mbox{}
  \caption{A comparison between our order-by-order fit (left) of $\eta\to3\pi^0$ (corresponding to the $\chi$PT approach \cite{BG} respecting the chiral symmetry)
  and the resummed-fit parametrization of this process with the parameters obtained by fitting the $O(p^6)$ \cite{BG} $\eta\to\pi^0\pi^+\pi^-$ amplitude at Adler zero
  (equivalent to the second numerical iteration of \cite{disperze}).}
\end{centering}
\end{figure}

Using our order-by-order fit of $\chi$PT data (respecting the chiral orders as $\chi$PT does), we plot the dependence on $s$ and $t$ of the Dalitz plot of
$\lvert\bar{M}(s,t,u)\rvert^2$ coming from NNLO $\chi$PT \cite{BG} on Fig.~5~(left).
To reproduce the dispersive methods \cite{disperze}, we fit our resummed-fit parameterized $\lvert M(s,t,u)\rvert^2$ of $\eta\to\pi^0\pi^+\pi^-$ to the chiral $O(p^6)$
results at the Adler zero ($s=\frac43M_\pi^2$), where these approaches believe the chiral expansion. Using these resumed-fit parameters, we can determine
Dalitz parametri\-zation for a neutral Dalitz plot on Fig.~5 (right). (This fit should be equivalent to the second numerical iteration in the Colangelo et al.~\cite{disperze}
approach -- modulo possible different inputs.) As we see on Fig.~5 the slope of Dalitz plot (corresponding to $\alpha$) has really changed the sign.
This can also affect the computation of the decay widths of these amplitudes and change the value of $R$.

\section{Summary}
We have a general method allowing us to construct a model-independent two-loop parametri\-zation of the amplitudes of interesting decay modes
$K\to3\pi$ and $\eta\to3\pi$, based only on analyticity, unitarity, crossing symmetry, relativistic invariance and chiral power-counting.
Our method can be implemented in a straightforward way in the presence of isospin breaking (i.e.~to describe the cusp).

We have verified that this description is fully compatible with the previous existing chiral two-loop
calculation for $\eta\to3\pi$ in the first order in isospin limit. It is however more general
and can help to understand also the discrepancy in the sign of slope in Dalitz parametrization between the standard chiral calculation and the previous
dispersive approaches. It indicates that the chiral expansion of the slope parameter $\alpha$ might have a convergence problem and that different
resummations of higher orders can have a bigger effect on it. This together with the prospects of the increase in statistics for $\eta \to 3 \pi$ decays in
future experiments ask for a more realistic description of the Dalitz plot structure which would go beyond the simple linear + quadratic
parametrization used so far in the literature.

\end{document}